\documentclass[12pt]{article}
\usepackage{amsfonts,pifont}
\usepackage{amsmath,chemarrow,chemarr}
\usepackage{amssymb}
\usepackage{amstext}
\usepackage{amsfonts}
\usepackage{graphicx,epsfig}
\usepackage{indentfirst}
\usepackage{hyperref}

\newtheorem{theorem}{Theorem}
\newtheorem{proposition}[theorem]{Proposition}
\makeatletter
\def\ExtendSymbol#1#2#3#4#5{\ext@arrow 0099{\arrowfill@#1#2#3}{#4}{#5}}
\def\RightExtendSymbol#1#2#3#4#5{\ext@arrow 0359{\arrowfill@#1#2#3}{#4}{#5}}
\def\LeftExtendSymbol#1#2#3#4#5{\ext@arrow 6095{\arrowfill@#1#2#3}{#4}{#5}}
\makeatother

\begin{document}
\baselineskip 20pt

\title{Classification of Arbitrary Multipartite Entangled States under Local Unitary Equivalence}
\author{Jun-Li Li$^1$ and
Cong-Feng Qiao$^{1,2}$\footnote{Corresponding author.}\\[0.5cm]
$^{1}$Department of Physics, Graduate University of Chinese Academy of Sciences \\
YuQuan Road 19A, Beijing 100049, China\\[0.2cm]
$^{2}$Theoretical Physics Center for Science Facilities (TPCSF),
CAS\\ YuQuan Road 19B, Beijing 100049, China}

\date{}
\maketitle

\begin{abstract}
We propose a practical method for finding the canonical forms of
arbitrary dimensional multipartite entangled states, either pure or
mixed. By extending the technique developed in one of our recent
works, the canonical forms for the mixed $N$-partite entangled states
are constructed where they have inherited local unitary symmetries
from their corresponding $N+1$ pure state counterparts. A systematic
scheme to express the local symmetries of the canonical form is also
presented, which provides a feasible way of verifying
the local unitary equivalence for two multipartite entangled states. \\


\end{abstract}

\section{Introduction}

Entanglement is one of the most important ingredients in quantum information
science; it gives impetus to the most extraordinary nonclassical
applications, such as teleportation and quantum computation, etc
\cite{quantum-computation}. It is now generally regarded that the
entanglement is a key physical resource in realizing many quantum information
tasks, thus the quantitative and qualitative study of entanglement become
more and more important. Though superficially entangled states show different
features---usually not all entangled states are functionally
independent---they may be intrinsically the same as far as the entanglement
property is concerned. Two entangled states are said to be equivalent in
implementing the same quantum information task if they can be obtained with
certainty from each other via local operation and classical communication
(LOCC). Theoretically, this LOCC equivalent class is such defined that within
the class any two quantum states are inter-convertible by local unitary (LU)
operators \cite{three-qubit}.

The characterization of bipartite entangled states under LU
equivalence can be well understood by using the singular value
(Schmidt) decomposition. However things turn out to be much more
complicated when the multipartite states are concerned. On one hand,
the characterization of multipartite entanglement can be done by
computing the local unitary invariants of the quantum states
\cite{LU-invariat-98}. Two entangled states are LU equivalent if they
have the same LU invariants; the relation between LU equivalence for
$n$-partite pure states and the $(n-1)$-partite mixed states has also
been observed and is used in constructing the local unitary
invariants \cite{Sergio-Wang-LU-tran, ZW-ShMF-LU-n-n+1}. The
parameters in local invariants grow dramatically as the number of
partite increases, and the problem of identifying and interpreting
independent invariants becomes very complicated \cite{Mark-Wooters}.
Recently a operationally meaningful measures has been introduced for
three-qubit entanglement \cite{Three-measures}. On the other hand,
one can chose certain bases and put the quantum states in some
canonical (standard) forms. Along this line, a canonical method was
proposed in Ref.\cite{Carteret-Sudbery}, though it was only given in
a set of constraints on the coefficients of the quantum state. Later,
this method was reformulated into a compact form
\cite{Verstraete-Moor}. By introducing the standard form for
multipartite states, Kraus proposed a general way to determine the LU
transformation between two LU equivalent $n$-qubit states
\cite{Kraus-PRL104-020504,Kraus-PRA82-032121}, however as the
dimension increases, degeneracy emerges between the identical
eigenvalues of the one partite reduced density matrix, and the
verification of LU equivalence becomes unpractical.

Recently, in \cite{HOSVD-LU-pure} we have proposed a practical method for
finding the canonical form of pure multipartite state by using the high order
singular value decompositions (HOSVDs) and  the local symmetry properties of
the tensor form quantum states. In this work, we generalize this method to
the mixed states where the canonical forms for arbitrary mixed multipartite
states are constructed. Also, we develop a systematic scheme to present the
local symmetries among the canonical forms, which provides a feasible way to
verify the LU equivalence of two quantum states regardless of the degeneracy
conditions.

The structure of the paper goes as follows. In section 2, we give a
brief introduction to the basic technique of HOSVD which is used in
our entanglement classification. In section 3, we reformulate the
entanglement classification for multipartite pure states under LU
equivalence in a neater form, and a practical classification method
for arbitrary multipartite mixed states is developed where the
canonical forms for mixed states are explicitly constructed. After
this complete classification of multipartite entangled states with
their canonical forms, in section 4 we develop a systematic scheme
for verifying the LU symmetry between two entangled states. In
section 5, practical examples of three- and four-qubit states are
given. Finally, some concluding remarks are presented in section 6.

\section{LU equivalence of multipartite quantum state}

A general $N$-partite entangled quantum state in the dimensions $I_1\times
I_2 \times \cdots \times I_N$ can be formulated in the following form:
\begin{eqnarray}
|\Psi\rangle=\sum_{i_1=1,i_2=1,...,i_N=1}^{I_1,I_2,\cdots,I_N}
\psi_{i_{1}i_{2}...i_{N}} |i_1\rangle|i_2\rangle...|i_N\rangle \; ,
\end{eqnarray}
where $\psi_{i_{1}i_{2}...i_{N}}\in \mathbb{C}$ are coefficients of the
quantum state in representative bases. Two quantum states are said to be LU
equivalent if they are inter-convertible by LU operators, which can be
schematically expressed as
\begin{eqnarray}
|\Psi'\rangle & = & \bigotimes_{i}^{N} U^{(i)} |\Psi\rangle
\nonumber
\\ & = &
\sum_{\substack{i_{1},i_{2},\cdots,i_{N} \\
i'_{1},i'_{2},\cdots,i'_{N}}} \psi_{i_{1}i_{2}\cdots i_{N}}
u^{(1)}_{i'_{1}i_1}|{i'_{1}\rangle u^{(2)}_{i'_{2}i_2}|i'_{2}\rangle
\cdots u^{(N)}_{i'_{N}i_N}|i'_{N}}\rangle \; \nonumber \\
& = & \sum_{i_{1},i_{2},\cdots,i_{N}} \psi'_{i_{1}i_{2}\cdots i_{N}}
|{i_{1},i_{2},\cdots,i_{N}}\rangle \; . \label{psi-equiv-coe}
\end{eqnarray}
Here, the coefficients $\psi_{i_{1}i_{2}...i_{N}}$ can also be treated as the
entries of a tensor $\Psi$ and hence the quantum states can be represented by
high order complex tensors. In the tensor form of $\Psi$, the unitary
operator $U^{(n)}$ acting on the $n$th partite is defined as
\begin{eqnarray}
(U^{(n)}\Psi )_{i_1i_2\cdots i_{n-1}i_n'i_{n+1\cdots i_N}} \equiv
\sum_{i_n} \psi_{i_1i_2\cdots i_{n-1}i_ni_{n+1}\cdots i_N}
u^{(n)}_{i'_{n}i_n} \; .
\end{eqnarray}

For bipartite pure state, the tensor $\Psi$ is a matrix $\Psi =
[\psi_{i_1i_2}] \in \mathbb{C}^{I_1\times I_2}$ (matrices with
complex numbers of $I_1$ rows and $I_2$ columns) where the
dimensions of the Hilbert space for each partite are $I_1$ and $I_2$
separately. The singular value decomposition (SVD) of the bipartite
state $\Psi$ of dimensions $I_1\times I_2$ reads
\begin{eqnarray}
\Lambda = U^{(1)} \cdot \Psi \cdot U^{(2)} =
\mathrm{diag}\{\lambda_1,\cdots,\lambda_{I}\} \; ,
\end{eqnarray}
where $\lambda_i\geq \lambda_j \geq 0, \forall\, i<j$, $I =
\mathrm{min}\{I_1,I_2\}$. $\Lambda$ has the following two
properties:
\begin{enumerate}
\item the singular values $\lambda_i, i\in \{1,\cdots, I\}$ of matrix
    $\Psi$ are uniquely defined.
\item $\Lambda$ is a diagonal matrix and uniquely defined (with
    prescribed order of the singular values).
\end{enumerate}
In this case, the singular values of the quantum state $\Psi$ readily
characterize its entanglement properties under LU equivalence. Two bipartite
quantum states are LU equivalent if, and only if, they have the same SVDs.

Here we introduce the technique which can be seen as the
generalization of SVD to high dimensional multipartite systems--the
HOSVD \cite{multi-singular-decom}. Let us define the matrix
unfolding of the tensor $\Psi\in \mathbb{C}^{I_1I_2 \cdots I_{N}} $
with $n$th index as
\begin{eqnarray}
\Psi_{(n)} \in \mathbb{C}^{I_n\times (I_{n+1}I_{n+2}\cdots
I_{N}I_1I_2\cdots I_{n-1})} \; .
\end{eqnarray}
Here $\Psi_{(n)}$ is a $I_n \times (I_{n+1}I_{n+2}\cdots
I_{N}I_1I_2\cdots I_{n-1})$ matrix. For example, the $2\times
3\times 4$ complex tensor $\Psi$, unfolding with the second and
third indexes, has the following forms:
\begin{eqnarray}
\Psi_{(2)} & = &
\begin{pmatrix}
\psi_{111} & \psi_{211} & \psi_{112} & \psi_{212} & \psi_{113} &
\psi_{213} & \psi_{114} & \psi_{214} \\ \psi_{121} & \psi_{221} &
\psi_{122} & \psi_{222} & \psi_{123} & \psi_{223} & \psi_{124} &
\psi_{224} \\ \psi_{131} & \psi_{231} & \psi_{132} & \psi_{232} &
\psi_{133} & \psi_{233} & \psi_{134} & \psi_{234}
\end{pmatrix} \; ,\nonumber \\
\Psi_{(3)} & = &
\begin{pmatrix}
\psi_{111} & \psi_{121} & \psi_{131} & \psi_{211} & \psi_{221} &
\psi_{231} \\ \psi_{112} & \psi_{122} & \psi_{132} & \psi_{212} &
\psi_{222} & \psi_{232} \\ \psi_{113} & \psi_{123} & \psi_{133} &
\psi_{213} & \psi_{223} & \psi_{233} \\ \psi_{114} & \psi_{124} &
\psi_{134} & \psi_{214} & \psi_{224} & \psi_{234}
\end{pmatrix} \; .
\end{eqnarray}
For arbitrary $N$-partite systems there exists a core tensor $\Omega$ for
each tensor $\Psi$,
\begin{eqnarray}
\Omega = U^{(1)}\otimes U^{(2)} \otimes \cdots \otimes U^{(N)}
\Psi\; . \label{Arrive-coretensor}
\end{eqnarray}
Here $\Omega$ is a same order tensor as $\Psi$ in the Hilbert space
of $I_1\times I_2\times \cdots \times I_N$. Any $N$-$1$ order tensor
$\Omega_{i_n=i}$ obtained by fixing the $n$th index to $i$, has the
following property:
\begin{eqnarray}
\langle \Omega_{i_n=i}, \Omega_{i_n=j} \rangle =
\delta_{\scriptstyle ij}\,\left(\sigma_{i}^{(n)}\right)^2\; ,
\end{eqnarray}
where $\sigma_{i}^{(n)}$ is called the $n$-mode singular value of
$\Psi$ and $ \sigma_{i}^{(n)} \geq \sigma_j^{(n)}\geq 0$, $\forall\;
i<j$ . The singular value $\sigma_i^{(n)}$ symbolizes the
Frobenius-norm $\sigma_{i}^{(n)} = ||\Omega_{i_n=i}|| \equiv
\sqrt{\langle\Omega_{i_n=i},\Omega_{i_n=i}\rangle}$, where the inner
product $\langle \mathcal {A}, \mathcal {B} \rangle \equiv
\sum_{i_{1}} \sum_{i_{2}} \cdot\cdot\cdot \sum_{i_{N}}
b_{i_1i_2...i_N} a^*_{i_1i_2...i_N}$ (see
\cite{multi-singular-decom} for details).

In the following we show how to get the core tensor by the LU
transformation $U^{(i)}, i\in \{1,\cdots, N\}$ in
Eq.(\ref{Arrive-coretensor}). A quantum state $\Omega$ with the same
dimension as $\Psi$ is LU equivalent to $\Psi$ if
\begin{eqnarray}
\Omega = U^{(1)}\otimes U^{(2)} \otimes \cdots \otimes U^{(N)} \Psi
\; , \label{Omega-Psi-fold}
\end{eqnarray}
where $U^{(i)}, i\in \{1,\cdots, N\}$ are unitary matrices. In the
matrix unfolding form, Eq.(\ref{Omega-Psi-fold}) can be rewritten as
\begin{eqnarray}
\Omega_{(n)} = U^{(n)} \cdot \Psi_{(n)} \cdot (U^{(n+1,\cdots,
n-1)})^{\mathrm{T}} \; . \label{Omega-Psi-unfold}
\end{eqnarray}
Here $U^{(n+1,\cdots n-1)} \equiv U^{(n+1)}\otimes U^{(n+2)} \otimes
\cdots \otimes U^{(N)}\otimes U^{(1)} \otimes \cdots \otimes
U^{(n-1)}$; $\Omega_{(n)}$ and $\Psi_{(n)}$ have the same dimensions:
$I_n$ rows and $(I_{n+1}\times I_{n+2} \cdots \times I_N\times I_1
\times \cdots \times I_{n-1})$ columns. Now consider the particular
case where $U^{(n)}$ is obtained from the singular value
decomposition of matrix $\Psi_{(n)}$, i.e.
\begin{eqnarray}
 U^{(n)}\cdot \Psi_{(n)} \cdot V^{(n)}=
\mathrm{diag} \{ \sigma_1^{(n)}, \sigma_{2}^{(n)}, \cdots,
\sigma_{I_n}^{(n)} \} \; , \label{n-singular-list}
\end{eqnarray}
where $U^{(n)}$ and $V^{(n)}$ are unitary matrix, and
$\sigma_{i}^{(n)}\geq \sigma_{j}^{(n)} \geq 0, \forall\, i<j$.
Eq.(\ref{Omega-Psi-unfold}) now can be written as
\begin{eqnarray}
\Omega_{(n)} = \mathrm{diag} \{ \sigma_1^{(n)}, \sigma_{2}^{(n)},
\cdots, \sigma_{I_n}^{(n)} \} \cdot V^{(n)\dag} \cdot
(U^{(n+1,\cdots, n-1)})^{\mathrm{T}} \; . \label{unfolding-singular}
\end{eqnarray}
It is clear that $\Omega_{(n)}$ has orthogonal rows
\begin{eqnarray}
\langle \Omega_{i_n=j}, \Omega_{i_n=k} \rangle = \delta_{jk}\,
\left(\sigma_{j}^{(n)}\right)^2\; . \label{row-orthognal-equ}
\end{eqnarray}
Eq.(\ref{row-orthognal-equ}) always holds if $U^{(n+1,\cdots, n-1)}$
is a unitary matrix. In the similar way we can obtain all the other
local unitary matrices $U^{(i)}, i\in \{1,2,\cdots, N\}$, and
eventually, the core tensors $\Omega$ of $\Psi$ can then be
constructed via Eq.(\ref{Omega-Psi-fold}).

From the construction of the core tensor, two of the important
properties of HOSVD (when compared to its bipartite counterpart) can
be concluded:
\begin{enumerate}
\item The $n$-mode singular values $\sigma_i^{(n)}, i\in \{1,\cdots, I_n\}, n\in \{1,\cdots,
N\}$, of $\Psi$ are uniquely defined.
\item If $\forall \, n\in \{1,\cdots, N\}$ the $n$-mode singular values
$\sigma_i^{(n)}$ are all distinct, then $\Omega'_{(n)} =
\Theta_{(n)} \Omega_{(n)}$ is also a HOSVD of $\Psi$ where
$\Theta_{(n)} = \mathrm{diag}\{ e^{i\theta_{1}^{(n)}}, \cdots,
e^{i\theta_{I_n}^{(n)}} \}$. Otherwise, let
$\sigma_1^{(n)}>\sigma_{2}^{(n)}> \cdots
> \sigma_{k_n}^{(n)}\geq 0$ denote the distinct $n$-mode singular
values of $\Omega_{(n)}$ with respective positive multiplicities
$\mu_1^{(n)}, \mu_2^{(n)}, \cdots,\mu_{k_n}^{(n)}$ where
$\sum_{i=1}^{k_n}\mu^{(n)}_{i} = I_n$. In this case,
\begin{eqnarray}
\Omega'_{(n)} = \left[\bigoplus_{i=1}^{k_n}u^{(n)}_{i} \right]\,
\Omega_{(n)} \equiv S^{(n)}\, \Omega_{(n)} \label{symm-relate-n}
\end{eqnarray}
is also a HOSVD of $\Psi$. Here $u^{(n)}_{i} \in
\mathbb{C}^{\mu^{(n)}_{i}\times \mu^{(n)}_{i}}$ are arbitrary
 $\mu^{(n)}_{i}\times \mu^{(n)}_{i}$ unitary matrices and constitute the diagonal blocks of $S^{(n)}$ which are conformal to
those $n$-mode singular values of $\Omega_{n}$ with multiplicity.
\end{enumerate}

From the second property it is clear that, unlike the bipartite
case, the core tensor $\Omega$ (HOSVD) of $\Psi$ is not uniquely
defined.

\section{Classification under local unitary equivalence}

In this section we propose entanglement classification scheme by
decomposing the LU equivalence of the quantum states into two
correlated problems: the HOSVD and LU symmetries. First, we give a
brief introduction to the entanglement classification of arbitrary
dimensional multipartite pure states which was first proposed in
\cite{HOSVD-LU-pure}, then we extend the method to the mixed states,
by which the canonical forms for entanglement classes of mixed states
under the LU equivalence can be constructed neatly.

\subsection{LU equivalence for multipartite pure states}

Due to the nonuniqueness of the core tensors, $\Omega$ can not be
identified as the entanglement classes of the quantum states. The
philosophy of our scheme in \cite{HOSVD-LU-pure} is that if we impose
this nonuniqueness as a local symmetry within the core tensors
themselves, then we can get the unique canonical forms. That is, if
we regard the core tensors $\Omega$ and $\Omega'$ which are related
by LU operators as the same entanglement class then the HOSVD can be
seen as the entanglement classification of the multipartite state
$\Psi$.

Suppose that the core tensor $\Omega$ have $k_n$ distinct $n$-mode
singular values $\sigma_{i}^{(n)}, i \in \{1, 2, \cdots, k_n\}$, each
with multiplicity of $\mu_{i}^{(n)}$ where
$\sum_{i=1}^{k_n}\mu^{(n)}_{i} = I_n$. Here we regard these
multiplicities as the degeneracies of the singular values which
corresponds to the case of nongeneric states of
\cite{Kraus-PRL104-020504}. From Eq.(\ref{symm-relate-n}) we can
infer that the LU symmetry which relates two core tensors takes the
following form
\begin{eqnarray}
S = \bigotimes_{n=1}^{N} \left[\bigoplus_{i=1}^{k_n}
u_{i}^{(n)}\right] \; .
\end{eqnarray}
The core tensors $\Omega'$ and $\Omega$ related by this symmetry now
can be written as
\begin{eqnarray}
\Omega' = S \, \Omega \; . \label{symmetry-between-coretensors}
\end{eqnarray}
Two different core tensors related by $S$ belong to the same
entanglement class. We can call such core tensor $\Omega$ of $\Psi$
associated with corresponding local symmetry $S$ the canonical form
of $\Psi$.

In order to see how this symmetry act on the core tensors we
introduce the technique of vectorization of the matrix. With each
matrix $A = [a_{ij}] \in \mathbb{C}^{I_1\times I_2}$, we can
associated it with a vector $\vec{A}$ defined by
\begin{eqnarray}
\vec{A} \equiv \left[a_{11}, \cdots, a_{I_11}, a_{12}, \cdots,
a_{I_1,2}, \cdots,  a_{1I_2}, \cdots, a_{I_1I_2}\right]^{\mathrm{T}}
\; . \label{vect-mat}
\end{eqnarray}
Two tensors $\Psi$ and $\Psi'$ of $I_1\times I_2\times \cdots I_N$
which are related by local operators $U^{(n)}, n\in \{1, 2, \cdots,
N\}$, can be expressed in the matrix unfolding form with the $n$th
index
\begin{eqnarray}
\Psi'_{(n)} = U^{(n)} \cdot \Psi_{(n)} \cdot ( U^{(n+1)}\otimes
U^{(n+2)} \otimes \cdots \otimes U^{(N)}\otimes U^{(1)} \otimes
\cdots \otimes U^{(n-1)})^{\mathrm{T}} \; ,
\label{matrixEq-to-vecEq}
\end{eqnarray}
With the convention of Eq.(\ref{vect-mat}), the matrix equation
Eq.(\ref{matrixEq-to-vecEq}), can be written as  (see
\cite{Horn-Johnson-2})
\begin{eqnarray}
U^{(n+1)}\otimes U^{(n+2)} \otimes \cdots \otimes U^{(N)}\otimes
U^{(1)} \otimes \cdots \otimes U^{(n-1)}\otimes U^{(n)}
\vec{\Psi}_{(n)} = \, \vec{\Psi}'_{(n)} \; .
\label{Eq-Core-tensor-Equi}
\end{eqnarray}
This can be seen as a unitary transformation of a $I_1\times
I_2\times \cdots \times I_N$ vector $\vec{\Psi}_{(n)}$ to
$\vec{\Psi}'_{(n)}$. On choosing $n=N$, we have the simple form of
Eq.(\ref{Eq-Core-tensor-Equi})
\begin{eqnarray}
U^{(1)} \otimes \cdots\otimes U^{(N)} \vec{\Psi}_{(N)} = \,
\vec{\Psi}'_{(N)} \; .
\end{eqnarray}
Here the symmetry between their core tensors,
Eq.(\ref{symmetry-between-coretensors}), can be similarly
represented as
\begin{eqnarray}
\vec{\Omega'}_{(N)} = S \, \vec{\Omega}_{(N)} \equiv
\bigotimes_{n=1}^{N} \left[\bigoplus_{i=1}^{k_n}
u_{i}^{(n)}\right] \vec{\Omega}_{(N)}  \; , \label{compact-sym}
\end{eqnarray}
where $u_{i}^{(n)}$ is a $\mu_{i}^{(n)}\times \mu_{i}^{(n)}$
unitary matrix and $\sum_{i=1}^{k_n} \mu_{i}^{(n)} = I_n$. In
the blocks diagonalized form, Eq.(\ref{compact-sym}) is
\begin{eqnarray}
\begin{pmatrix}
u^{(1)}_1\otimes \cdots \otimes u^{(N)}_1 & 0 & \cdots &  0 \\
0 & u^{(1)}_1\otimes \cdots \otimes u^{(N)}_2 & \cdots &   0 \\
\vdots & \vdots & \ddots & \vdots \\
0 & 0 & \cdots & u^{(1)}_{k_1}\otimes \cdots \otimes u^{(N)}_{k_N}
\end{pmatrix} \cdot
\vec{\Omega}_{(N)} =
\vec{\Omega'}_{(N)} \; .
\label{mat-vec-decompose}
\end{eqnarray}
We can set $u^{(n)}_{j} = e^{i \theta_{j}^{(n)}}$ if the
multiplicity $\mu^{(n)}_{j} = 1$. In all, we have the following
theorem which has been state in \cite{HOSVD-LU-pure}
\begin{theorem}
The core tensors $\Omega$ associated with the local symmetry group
$S$ is the canonical form of the multipartite pure state and is the
entanglement class under LU equivalence.
\end{theorem}

\begin{figure}[t]\centering
\scalebox{0.6}{\includegraphics{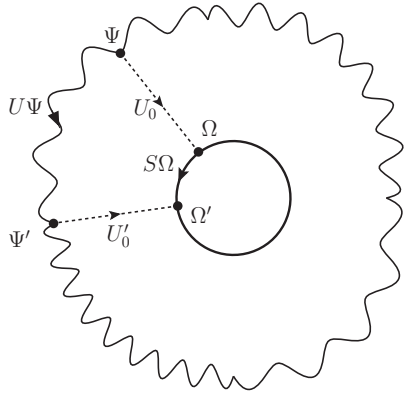}} \caption{ The
wiggled line represents $U\Psi$ and it forms an orbit with irregular
shapes; the circle line represents $S\Omega$ and it forms
well-structured orbit. $\Omega=U_0\Psi$ and $\Omega'=U_o'\Psi'$ both
are the core tensors on the $S\Omega$ orbit.} \label{fig-map-HOSVD}
\end{figure}

From this theorem, we can form a more general point of view for the
equivalent entanglement class. Any subset of the quantum states in
$I_1\times I_2\times \cdots \times I_N$, i.e., $\{\psi\}$, associated
with its local unitary transformation group $U=\bigotimes_iU^{(i)}$
where $\{\psi| U\psi \in \{\psi\}\}$ can be regarded as a unique
representation of entanglement class. Let $\{\Psi, U\}$ be such
subset of the quantum states associated with its LU symmetry, then it
would be intrinsically the same as the classification with the HOSVD
$\Omega$ and its LU symmetry group $S$, i.e., $\{\Omega, S\}$.
However the core tensors $\Omega$ have nice properties of much simple
form of local symmetries $S$, that is
\begin{eqnarray}
S = \bigotimes_{n=1}^{N}
\left[\bigoplus_{j=1}^{k_n} u_{j}^{(n)}\right]\; .
\end{eqnarray}
 The procedure of our entanglement classification can be
formulated as (see Fig.(\ref{fig-map-HOSVD}))
\begin{eqnarray}
& & \Psi' = U \Psi \nonumber \\ & \Rightarrow & U_0' \Psi'  =
(U_0' U U_0^{\dag})\cdot \left( U_0 \Psi \right) \nonumber \\
\ & \Rightarrow & \Omega' = S \cdot \Omega \; ,
\label{conjugate-S-U}
\end{eqnarray}
where $U,\ S,\ U_0,\ U_0'\in \bigotimes_i U^{(i)}$. In
Eq.(\ref{conjugate-S-U}), to some extent, $S$ can be seen as a
conjugate class of $U$ under a particular local unitary
transformation $U_0'$ and $U_0^{\dag}$. In the special case that all
the singular values are distinct for each partite, the symmetry
becomes
\begin{eqnarray}
S = \bigotimes_{n=1}^{N} \left[\bigoplus_{j=1}^{I_n}
e^{i\theta_{j}^{(n)}}\right]\;,
\end{eqnarray}
which is just the conjugate class of $\bigotimes_{i=1}^{N}U^{(i)}$
under $\bigotimes_{i=1}^{N}U^{(i)}$.

From the quantum state point of view, tensor $\Omega$ now is
decomposed into several invariant subtensors (denoted it by $\omega$)
of the Hilbert space of $I_1\times I_2\times \cdots \times I_N$ under
the transformation $S$. The dimensions of these subtensors conformal
to the direct-summed subgroups of $S$ in
Eq.(\ref{mat-vec-decompose}). For example
Eq.(\ref{mat-vec-decompose}) can be written as
\begin{eqnarray}
\begin{pmatrix}
u^{(1)}_1\otimes \cdots \otimes u^{(N)}_1 & 0 & \cdots &  0 \\
0 & u^{(1)}_1\otimes \cdots \otimes u^{(N)}_2 & \cdots &   0 \\
\vdots & \vdots & \ddots & \vdots \\
0 & 0 & \cdots & u^{(1)}_{k_1}\otimes \cdots \otimes u^{(N)}_{k_N}
\end{pmatrix} \cdot
\begin{pmatrix}
\vec{\omega}_{r_1} \\ \vec{\omega}_{r_2} \\ \vdots \\
\vec{\omega}_{r_m}
\end{pmatrix} =
\begin{pmatrix}
\vec{\omega}'_{r_1} \\ \vec{\omega}'_{r_2} \\ \vdots \\
\vec{\omega}'_{r_m}
\end{pmatrix} \; , \label{detailed-symmetry-equation}
\end{eqnarray}
where $\vec{\omega}_{r_i}$ and $\vec{\omega}'_{r_i}$ are the
segments of the column vectors $\vec{\Omega}_{(N)}$ and
$\vec{\Omega}'_{(N)}$ with the dimension conformal to the diagonal
blocks $u^{(1)}_{i_1}\otimes \cdots \otimes u^{(N)}_{i_n}$.
$\vec{\omega}$s are just the vector forms of the subtensors
$\omega$.

\subsection{LU equivalence for multipartite mixed states}

The classification of the entanglement for mixed states is generally
believed to be more complicated than pure states in many cases.
However in the case of LU equivalence, it has been noticed that
$n$-partite pure state is related to its $(n-1)$-partite mixed state
\cite{ZW-ShMF-LU-n-n+1}. Here we generalize our entanglement
classification method developed for multipartite pure states to the
case of arbitrary dimensional $N$-partite mixed states.

Consider a mixed $N$-partite quantum state $\rho$ which is generally
expressed as
\begin{eqnarray}
\rho = \sum_{i}p_i^2\, |\psi_i\rangle \langle \psi_i| \; ,
\end{eqnarray}
where $\sum_{i} p_i^2 = 1$,  $p_i \in \mathbb{R}^{+}$,
$|\psi_i\rangle$ are $N$ partite pure states. We add an additional
$0$th partite to the original $N$-partite mixed state $\rho$ and
formulate an $N+1$ pure quantum state in the following form
\begin{eqnarray}
\Psi_{0} = \sum_{i} p_i |i\rangle |\psi_i\rangle \; ,
\end{eqnarray}
where $|i\rangle$ are the bases of 0th partite. For this quantum
state, we have the following fact:
\begin{eqnarray}
\mathrm{Tr}_{0}\left[|\Psi_0\rangle \langle \Psi_0| \right] & = &
\sum_{n,i,j} p_ip_j \langle n |i\rangle |\psi_i\rangle \langle
\psi_j|\langle j| n\rangle \nonumber \\ & = & \sum_{n,i,j} p_ip_j
\langle j| n\rangle \langle n |i\rangle |\psi_i\rangle \langle
\psi_j| \nonumber
\\ & = & \sum_{i} p_i^2 |\psi_i\rangle \langle \psi_i| = \rho \; .
\end{eqnarray}
Further we have, if $\Psi'_0 = U^{(0)}\otimes E^{(1)}\otimes
\cdots\otimes E^{(N)} \Psi_0 \equiv U^{(0)}\, \Psi_0$, where $E$ is
unit matrix, then
\begin{eqnarray}
& & \mathrm{Tr}_{0}\left[|\Psi'_0\rangle \langle \Psi'_0| \right] =
\mathrm{Tr}_{0}\left[U^{(0)}|\Psi_0\rangle \langle \Psi_0
|U^{(0)\dag} \right] \nonumber \\ & = & \sum_{n,i,j} p_ip_j \langle
n|U^{(0)} |i\rangle |\psi_i\rangle \langle \psi_j |\langle
j|U^{(0)\dag}|n\rangle  \nonumber \\ & = & \sum_{n,i,j} p_ip_j
\langle j|U^{(0)\dag}|n\rangle \langle n|U^{(0)} |i\rangle
|\psi_i\rangle \langle \psi_j |\nonumber
\\ & = & \rho =
\mathrm{Tr}_{0}\left[|\Psi_0\rangle \langle \Psi_0|\right] \; . \label{Tr0-psi}
\end{eqnarray}
From the above two facts we can state that the following relation
\begin{eqnarray}
\rho = \sum_{i=1}^{r}p_i^2|\psi_i\rangle\langle \psi_i|
\xrightleftharpoons[+0 ]{\mathrm{Tr}_0}  \; \Psi_0 = \sum_{i=1}^{r}
p_i|i\rangle |\psi_i\rangle
\end{eqnarray}
forms a bijection between $\rho$ and $\Psi_0$. Define this bijection
as a map between $\Psi_0$ and $\rho$, we have the following
proposition
\begin{proposition}
An arbitrary dimensional mixed $N$-partite state $\rho'$ is LU
equivalent to $\rho$, i.e.,
\begin{eqnarray}
\rho' = U^{(1)}\otimes U^{(2)}\otimes \cdots \otimes U^{(N)} \rho\,
U^{(1)\dag}\otimes U^{(2)\dag}\otimes \cdots \otimes U^{(N)\dag}
\end{eqnarray}
if and only if its pure state counterpart $\Psi'_0$ is LU equivalent
to $\Psi_0$, i.e.,
\begin{eqnarray}
\Psi'_0 = U^{(0)}\otimes U^{(1)}\otimes \cdots \otimes U^{(N)}
\Psi_0 \; .
\end{eqnarray}
\end{proposition}

\noindent Proof: First, if
\begin{eqnarray}
\rho' & = & U^{(1)}\otimes U^{(2)}\otimes \cdots \otimes U^{(N)}
\rho\, U^{(1)\dag}\otimes U^{(2)\dag}\otimes \cdots \otimes
U^{(N)\dag} \nonumber \\  & = & \sum_{i=1}^{r} p_i^2 U^{(1)}\otimes
U^{(2)}\otimes \cdots \otimes U^{(N)} |\psi_i\rangle \langle \psi_i|
U^{(1)\dag}\otimes U^{(2)\dag}\otimes \cdots \otimes U^{(N)\dag}
\nonumber \\ & = & \sum_{i=1}^{r} p^2_i |\psi_i'\rangle \langle
\psi_i'| \; ,
\end{eqnarray}
where $|\psi_i'\rangle = U^{(1)}\otimes U^{(2)}\otimes \cdots
\otimes U^{(N)} |\psi_i\rangle$, then $\Psi'_0$ correspond to
$\rho'$ is
\begin{eqnarray}
\Psi'_0 & = & \sum_{j=1}^{r} p_j|j\rangle |\psi_j'\rangle =
\sum_{j=1}^{r} p_j |j\rangle U^{(1)}\otimes U^{(2)}\otimes \cdots
\otimes U^{(N)} |\psi_j\rangle \nonumber \\ & = & E^{(0)} \otimes
U^{(1)}\otimes U^{(2)}\otimes \cdots \otimes U^{(N)} \sum_{j=1}^{r}
p_j |j\rangle |\psi_j\rangle \nonumber \\ & = & E^{(0)} \otimes
U^{(1)}\otimes U^{(2)}\otimes \cdots \otimes U^{(N)}\Psi_0 \; .
\end{eqnarray}
That is, $\Psi'_0$ is LU equivalent to $\Psi_0$.

Second if $\Psi'_0 = U^{(0)}\otimes \cdots \otimes U^{(N)}\Psi_0$,
then
\begin{eqnarray}
\rho' & = & \mathrm{Tr}_0\left[ |\Psi'_0\rangle \langle \Psi'_0|
\right] \nonumber \\ & = & \mathrm{Tr}_0\left[ \sum_{i=1,j=1}^{r}
p_ip_jU^{(0)}|i\rangle U^{(1)}\otimes \cdots \otimes U^{(N)}
|\psi_i\rangle \langle \psi_j| U^{(1)\dag}\otimes \cdots \otimes
U^{(N)\dag} \langle j| U^{(0)\dag} \right] \nonumber \\ & = &
\mathrm{Tr}_0\left[ \sum_{i=1,j=1}^{r} p_ip_j |i\rangle
U^{(1)}\otimes \cdots \otimes U^{(N)} |\psi_i\rangle \langle \psi_j|
U^{(1)\dag}\otimes \cdots \otimes U^{(N)\dag} \langle j|\right]
\nonumber \\ & =& U^{(1)}\otimes \cdots \otimes U^{(N)} \left[
\sum_{i=1}^{r} p_i^2  |\psi_i\rangle \langle \psi_i| \right]
U^{(1)\dag}\otimes \cdots \otimes U^{(N)\dag} \nonumber \\ & = &
U^{(1)}\otimes \cdots \otimes U^{(N)} \rho\, U^{(1)\dag}\otimes
\cdots \otimes U^{(N)\dag} \; .
\end{eqnarray}
Here, we have used the fact of Eq.(\ref{Tr0-psi}). That is $\rho'$
is LU equivalent to $\rho$. QED.

We can now conclude that: if $\rho'$ is LU equivalent to $\rho$ then
their corresponding pure states $\Psi'_0$ and $\Psi_0$ can be
related by LU operators; if $\Psi'$ is LU equivalent to $\Psi_0$
then their reduced matrices $\rho'$ and $\rho$ are LU equivalent. We
may construct the core tensor $\Omega_0$ from $\Psi_0$, then we
trace out the $0$th partite from the core tensor $\Omega_0$ and
obtain the canonical form for $\rho$, that is
\begin{eqnarray}
\Upsilon = \mathrm{Tr}_0 \left[|\Omega_0\rangle \langle
\Omega_0|\right] \; .
\end{eqnarray}
\begin{theorem}
The canonical form $\Upsilon$ is of the entanglement class of mixed
state $\rho$ up to a local symmetry inherit from $\Omega_0$.
\end{theorem}

This method provides a simple way to construct the canonical form for
the mixed $N$-partite state $\rho$: first construct the $N$+1 partite
pure state $\Psi_0$ from $\rho$; then compute the core tensor
$\Omega_0$ of $\Psi_0$; finally we arrive at the canonical form by
tracing out the 0th partite $\Upsilon =\mathrm{Tr}_0|\Omega_0\rangle
\langle \Omega_0|$.

\section{The local symmetries of the canonical form}

We have constructed the canonical forms for both pure and mixed
multipartite states. In all, the construction of the canonical forms
will result in a general form of Eq.(\ref{mat-vec-decompose})
whether the state is pure or not. With this direct summed forms of
the symmetries, in this section we develop a practical scheme to
verify the LU equivalence of two quantum states which have the same
singular values and same degeneracies for each partite.

\subsection{A general form of the local unitary symmetry}

We start from a general case, that is we have $k_n$ distinct
$n$-mode singular values $\sigma_{i_n}^{(n)}, i_n \in \{1, 2,
\cdots, k_n\}$, each with multiplicities of $\mu_{i_n}^{(n)}$ where
$\sum_{i_n=1}^{k_n}\mu^{(n)}_{i_n} = I_n$. Define the $n$-mode
singular value vector $\vec{\sigma}^{(n)}$ for the matrix unfolding
form of $\Omega_{(n)}$
\begin{eqnarray}
\vec{\sigma}^{(n)} & \equiv  & \{\; \underbrace{\sigma_1^{(n)} ,
\sigma_1^{(n)} , \cdots , \sigma_1^{(n)}}_{\mu^{(n)}_1} ,\;
\underbrace{\sigma_2^{(n)}, \sigma_2^{(n)}, \cdots ,
\sigma_2^{(n)}}_{\mu^{(n)}_2} , \nonumber \\ & & \hspace{2.5cm}
\vdots \nonumber \\ & & \hspace{0.35cm}
\underbrace{\sigma_{k_n}^{(n)} , \sigma_{k_n}^{(n)} , \cdots ,
\sigma_{k_n}^{(n)}}_{\mu^{(n)}_{k_n}}\; \}^{\mathrm{T}} \; ,
\end{eqnarray}
where $\sigma_i^{(n)} > \sigma_j^{(n)} \geq 0, \forall\; i<j$. The
local symmetry corresponding to this partite is
\begin{eqnarray}
S^{(n)} \equiv \begin{pmatrix}
u^{(n)}_{1} &   &   \\  & \ddots &   \\
 &   & u^{(n)}_{k_n}
\end{pmatrix} \; ,
\end{eqnarray}
where $u^{(n)}_{i}, i \in \{ 1, \cdots , k_n\}$ are unitary with
the dimension of $\mu_{i}^{(n)} \times \mu_{i}^{(n)}$.

The total local unitary symmetry $ S = \bigotimes_i S^{(i)}$ of the
core tensors $\Omega$ is
\begin{eqnarray}
\begin{pmatrix}
u^{(1)}_{1} &   &   \\  & \ddots &   \\
 &   & u^{(1)}_{k_1}
\end{pmatrix}  \otimes \cdots \otimes
\begin{pmatrix}
u^{(N)}_{1} &   &   \\  & \ddots &   \\
  &   & u^{(N)}_{k_N}
\end{pmatrix} \cdot \vec{\Omega}_{(N)} = \vec{\Omega}'_{(N)}\; ,
\end{eqnarray}
which is just Eq.(\ref{mat-vec-decompose}). Define the ``singular
value matrix'' $\Sigma$ of the core tensor
\begin{eqnarray}
\Sigma \equiv \{\vec{\sigma}^{(1)}, \vec{\sigma}^{(2)}, \cdots,
\vec{\sigma}^{(N)}\}\; ,
\end{eqnarray}
where it is uniquely defined according to the properties of HOSVD.
Quantum states with different singular value matrices are apparently
LU inequivalent. Then the core tensors which have the same singular
value matrix belong to the same entanglement class if and only if
they satisfy Eq.(\ref{detailed-symmetry-equation}). In
Eq.(\ref{detailed-symmetry-equation}), the verification of the LU
equivalence of two core tensors turns to finding the solutions of
the following equation groups with varying $r$
\begin{eqnarray}
u^{(1)}_{i_1} \otimes u^{(2)}_{i_2}\otimes \cdots \otimes
u^{(N)}_{i_{N}} \vec{\omega}_{r} = \vec{\omega}_{r} \; .
\end{eqnarray}
This can be seen as a fine-grained LU classification problem of the
subtensor $\omega_{r}$. Thus we can pick the sub tensor $\omega_{r}$
out of the tensor $\Omega$ and do the HOSVD to it recursively.

As the fine-grained process goes, the recursive procedure will
terminated at two conditions: 1, the singular values are all distinct
for all the partite; 2, the singular values are all the same for all
the partite.

For the first case, if $\forall\, i \in \{1,\cdots, k_n\}$ and
$\forall\, n\in \{1,\cdots, N\}$, the singular value multiplicity
$\mu_{i}^{(n)} = 1$, then $k_n = I_n$ and
\begin{eqnarray}
U^{(n)} =
\begin{pmatrix}
\exp(i\theta_{1}^{(n)}) & 0 & \cdots  & 0 \\ 0  &
\exp(i\theta^{(n)}_{2}) & \cdots & 0
\\  \vdots &  \vdots  & \ddots &  \vdots \\  0 & 0 & \cdots  & \exp(i\theta^{(n)}_{I_n})
\end{pmatrix} \; ,
\end{eqnarray}
Eq.(\ref{detailed-symmetry-equation}) turns to
\begin{eqnarray}
\exp[i(\theta^{(1)}_{i_1}+\theta^{(2)}_{i_2}+\cdots +
\theta^{(N)}_{i_N})] \omega_{i_1i_2\cdots i_N} =
\omega'_{i_1i_2\cdots i_N} \; . \label{all-distinct}
\end{eqnarray}
Here we write $\omega_{i_1i_2\cdots i_N}$ instead of $\vec{\omega}$
because $\omega_{i_1i_2\cdots i_N}$ now is a complex number. A $\log$
operation on Eq.(\ref{all-distinct}) would result in a linear
functional group
\begin{eqnarray}
\theta^{(1)}_{i_1}+\theta^{(2)}_{i_2}+\cdots + \theta^{(N)}_{i_N}
 = -i \log\left[\frac{\omega'_{i_1i_2\cdots i_N}}{\omega_{i_1i_2\cdots i_N}}\right] \; .
\end{eqnarray}
These are $I_1\times I_2\times \cdots \times I_N$ linear equations
for $I_1 + I_2 + \cdots + I_N$ phase variables and it can be verified
immediately whether they have consistent solutions. The quantum
states is LU equivalent if, and only if, there is at leat one
solution to this linear equation group.

\subsection{A completely degenerate state for all the partite}

In the completely degenerate state, the reduced density matrix for
each partite is proportional to unit matrix. Consider a arbitrary
$N$-partite pure state with dimension of $I_1\times I_2\times \cdots
I_N$. The complete degenerate state is that $\forall\, n\in \{1,\,
2,\,\cdots, N\}$
\begin{eqnarray}
\rho_{n} =  \mathrm{Tr}_{\neg\, n}\left[ | \Omega \rangle \langle
\Omega | \right] = \frac{1}{I_n} \, E \; .
\end{eqnarray}
The core tensor has the following form
\begin{eqnarray}
\Omega & = & \sum_{i_n}|i_n\rangle \sum_{\neg \, i_n}
\omega_{i_{1}i_{2}\cdots i_{n-1}i_n i_{n+1}\cdots i_{N}}
|{i_{1}i_{2}\cdots i_{n-1}i_{n+1}\cdots i_{N}}\rangle \nonumber \\ &
= & \sum_{i_n}|i_n\rangle |\omega^{(i_n)}_{\neg\, n}\rangle\; , \;
n\in \{1,\cdots, N\} \; ,
\end{eqnarray}
where $\langle \omega^{(i_n')}|\omega^{(i_n)}\rangle =
\frac{1}{I_n}\delta_{i_ni_n'}$. The local symmetry $S$ takes the
following form
\begin{eqnarray}
\vec{\Omega}'_{(N)} = S\cdot \vec{\Omega}_{(N)} =\bigotimes_{n}
U^{(n)} \cdot \vec{\Omega}_{(N)}  \; . \label{total-symmetry}
\end{eqnarray}
An arbitrary unitary matrix is unitarily equivalent to a diagonal
matrix, that is
\begin{eqnarray}
U^{(n)} = X^{(n)\dag} \cdot \Phi^{(n)} \cdot X^{(n)} \; ,
\end{eqnarray}
where $X^{(n)}$ is unitary matrix and $\Phi^{(n)} =
\text{diag}\{e^{i\phi^{(n)}_1},\cdots, e^{i\phi^{(n)}_{I_n}}\}$ is
the conjugate class of $U^{(n)}$. Eq.(\ref{total-symmetry}) now
turns to
\begin{eqnarray}
\bigotimes_{n} \Phi^{(n)}\cdot  \bigotimes_{n} X^{(n)} \cdot
\vec{\Omega} = \bigotimes_{n} X^{(n)} \cdot \vec{\Omega}'\; .
\label{U-equ-decompose}
\end{eqnarray}
The Eq.(\ref{U-equ-decompose}) corresponds to $I_1\times I_2\times
\cdots \times I_N$ homogeneous equations, which in the detailed form
one of the of these equations of Eq.(\ref{U-equ-decompose}) looks
like
\begin{eqnarray}
\sum_{i_1\cdots i_N} x_{j_1i_1}^{(1)} x^{(2)}_{j_2i_2} \cdots
x^{(N)}_{j_Ni_N} \cdot (e^{i(\phi^{(1)}_{j_1}+\phi^{(2)}_{j_2} +
\cdots \phi^{(N)}_{j_N})}\, \omega_{i_1i_2\cdots i_N} -
\omega'_{i_1i_2\cdots i_N}) =0 \; . \label{over-define-equ}
\end{eqnarray}
Here we represent $x_{ij}$ as the elements of matrix $X$. This is a
typical equation group of $I_1\times I_2\times \cdots \times I_N$
equations for $I_1^2+I_2^2+ \cdots + I_N^2$ complex parameters
$x^{(n)}_{i_ni_n'}$ (note we first solve the parameters
$x_{ij}^{(n)}$ then impose the unitary condition on the matrix
$X^{(n)}$).

For this kind of nonlinear equations there exist simple tool called
``linearization'' or ``relinearization'' \cite{Kipnis-Shamir,
Courtois-Klimov-Patarin-Shamir}. The key algorithm rely on the fact
that for $N>2$ multipartite quantum states, when the dimensional or
number of partite increases, the number of equations grows much more
quickly than the number of the parameters. Generally
Eq.(\ref{over-define-equ}) would turn out to be an over-defined
system of equations which mean that there are more equations than
unknow parameters.

The linearization technique goes as follows. Regard each monomial of
the matrix elements as a individual variable
\begin{eqnarray}
\nu_{i_1i_2\cdots i_N,i'_1i'_2\cdots i'_N} = x_{i_1i'_1}^{(1)}
x^{(2)}_{i_2i'_2} \cdots x^{(N)}_{i_Ni'_N} \; ,
\label{linearization-assign}
\end{eqnarray}
then there will be $(I_1\times I_2\times \cdots \times I_N)^2$ such
variables $\nu$. Eq.(\ref{over-define-equ}) now can be written as
\begin{eqnarray}
\left[ \sum_{i_1\cdots i_N} \nu_{j_1j_2\cdots j_N,i_1i_2\cdots i_N}
(e^{i(\phi^{(1)}_{j_1}+\phi^{(2)}_{j_2} + \cdots
\phi^{(N)}_{j_N})}\, \omega_{i_1i_2\cdots i_N} -
\omega'_{i_1i_2\cdots i_N}) \right] =0 \; .\label{over-define-equ-1}
\end{eqnarray}
For the sake of simplicity we use the convention that $\mathbf{i}$
represents the value of the bit string $(i_1i_2\cdots i_N)$, i.e.,
$\mathbf{i}=1=(11\cdots 1)$ and $\mathbf{i}=2=(11\cdots 2)$, etc.
Define $\omega_{\mathbf{j}\mathbf{i}} \equiv
e^{i(\phi^{(1)}_{j_1}+\phi^{(2)}_{j_2} + \cdots \phi^{(N)}_{j_N})}\,
\omega_\mathbf{i} - \omega'_\mathbf{i}$ where $\mathbf{j} = (j_1j_2
\cdots j_N)$. Eq.(\ref{over-define-equ-1}) can be reformulated as
\begin{eqnarray}
\omega_{\mathbf{j}\mathbf{i}} \cdot \nu_{\mathbf{j}\mathbf{i}} = 0\;
,
\end{eqnarray}
where the dot means the summation over $\mathbf{i}$. Taking a
$2\times 2\times 2$ system as an example, we have
\begin{eqnarray}
& & \omega_{\mathbf{j}1} \nu_{\mathbf{j}1} + \omega_{\mathbf{j}2}
\nu_{\mathbf{j}2} + \omega_{\mathbf{j}3} \nu_{\mathbf{j}3} +
\omega_{\mathbf{j}4} \nu_{\mathbf{j}4} + \nonumber \\ & &
\omega_{\mathbf{j}5} \nu_{\mathbf{j}5} +\omega_{\mathbf{j}6}
\nu_{\mathbf{j}6} + \omega_{\mathbf{j}7} \nu_{\mathbf{j}7} +
\omega_{\mathbf{j}8} \nu_{\mathbf{j}8} = 0 \; .
\end{eqnarray}
There are 8 such equations for $\mathbf{j}$ runs from $1$ to $8$.
The solution can be expressed as
\begin{eqnarray}
\begin{pmatrix}
\nu_{\mathbf{j}1} \\ \nu_{\mathbf{j}2} \\ \nu_{\mathbf{j}3} \\ \vdots \\
\nu_{\mathbf{j}8}
\end{pmatrix} = c_{\mathbf{j}2}
\begin{pmatrix}
-\frac{\omega_{\mathbf{j}2}}{\omega_{\mathbf{j}1}} \\ 1 \\ 0 \\ \vdots \\
0
\end{pmatrix} + c_{\mathbf{j}3}
\begin{pmatrix}
-\frac{\omega_{\mathbf{j}3}}{\omega_{\mathbf{j}1}} \\ 0 \\ 1 \\ \vdots \\
0
\end{pmatrix} + \cdots + c_{\mathbf{j}8}
\begin{pmatrix}
-\frac{\omega_{\mathbf{j}8}}{\omega_{\mathbf{j}1}} \\ 0 \\ 0 \\ \vdots \\
1
\end{pmatrix} \; ,\label{underdefine-equation}
\end{eqnarray}
where $c_{\mathbf{j}\mathbf{i}}$ are new parameters. Clearly,
Eq.(\ref{underdefine-equation}) is a under defined equation group for
parameters $\nu_\mathbf{ji}$. However, there are additional equations
between the products of $\nu_{\mathbf{ji}}$\, s, i.e.
\begin{eqnarray}
& & \nu_{i_1 \cdots i_n \cdots i_N, j_1 \cdots j_n \cdots j_N}
\nu_{i'_1 \cdots i'_n\cdots i'_N,j'_1 \cdots j'_n\cdots j'_N} \nonumber \\
& = & \nu_{i_1 \cdots i'_n\cdots i_N,j_1 \cdots j'_n\cdots j_N}
\nu_{i'_1 \cdots i_n\cdots i'_N,j'_1 \cdots j_n\cdots j'_N} \; .
\label{relinerization-param}
\end{eqnarray}
This relation is inherited from Eq.(\ref{linearization-assign}) as
\begin{eqnarray}
& & x_{i_1j_1}^{(1)} \cdots x^{(n)}_{i_nj_n} \cdots x^{(N)}_{i_Nj_N}
\cdot x_{i'_1j'_1}^{(1)} \cdots x^{(n)}_{i'_nj'_n} \cdots
x^{(N)}_{i'_Nj'_N} \nonumber \\ & = & x_{i_1j_1}^{(1)} \cdots
x^{(n)}_{i'_nj'_n} \cdots x^{(N)}_{i_Nj_N} \cdot x_{i'_1j'_1}^{(1)}
\cdots x^{(n)}_{i_nj_n} \cdots x^{(N)}_{i'_Nj'_N} \; .
\label{relinerization-commut-all}
\end{eqnarray}
For example in $2\times 2\times 2$ system we have $\nu_{111,
111}\nu_{111, 122}=\nu_{111, 121}\nu_{111, 112}$ or simply
$\nu_{11}\nu_{14} = \nu_{13}\nu_{12}$. This imposes an additional
equation between parameters $\nu_{\mathbf{ji}}$, and can also be
viewed as an equation in the (smaller number of) parameters
$c_{\mathbf{ji}}$ expressing them. The new system of equations can
be derived from all the possible relations of the type of
Eq.(\ref{relinerization-param}). In solving the equations on
$c_{\mathbf{ji}}$ we can use the linearization method recursively.

Here we give a explicit formula for how many constrains of
Eq.(\ref{relinerization-param}) there will be. As there are
$(I_1\times I_2\times \cdots \times I_N)^2$ matrix elements, we can
get $(I_1\times I_2\times \cdots \times I_N)^2$ new parameters
$\nu_{\mathbf{ji}}$. If we multiply $m$ times of
$\nu_{\mathbf{ji}}$, i.e.,
\begin{eqnarray}
\underbrace{\nu_{\mathbf{ji}}\,\cdots \, \nu_{\mathbf{j'i'}}}_{m} \;
,
\end{eqnarray}
we will have
\begin{eqnarray}
\mathrm{C}_{(I_1\times I_2\times \cdots \times I_N)^2+m-1}^{\,m}
\label{num-of-prod}
\end{eqnarray}
different productions. On the contrary, according to the productions
of $x_{ij}^{(n)}$, the actual number of different productions is
only
\begin{eqnarray}
\prod_{i=1}^{N} \mathrm{C}_{I_i^2+m-1}^{\,m} \; ,
\label{num-of-para}
\end{eqnarray}
which is much less than the number of equations (here
$\mathrm{C}_{n}^{\,l} = \frac{n!}{l!(n-l)!}$). The number of
Eq.(\ref{num-of-prod}) is greater than that of
Eq.(\ref{num-of-para}) when $m>1$. For the case of $2\times 2\times
2$ and $m=2$ we have
\begin{eqnarray}
\mathrm{C}_{(2\times 2 \times 2)^2+2-1}^{\,2} = 2080 \; , \; \left(
\mathrm{C}_{2^2+2-1}^{\,2} \right)^3= 1000 \; ,
\end{eqnarray}
which means that we have $2080$ different
$\nu_{\mathbf{ji}}\nu_{\mathbf{j}'\mathbf{i}'}$s, but only $1000$
are independent. A considerably large amount of constraint equations
like Eq.(\ref{relinerization-param}) are obtained.

There are actually many other methods and algorithms which are
applicable in finding the local unitary solutions that connect the
two entangled states, i.e., Gr\"obner basis
\cite{Grobner-basis-Buchberger}, FXL algorithm
\cite{Courtois-Klimov-Patarin-Shamir} etc. The application of them
lead to a connection between the local unitary transformational
matrices and the well-developed theory of algebraic varieties, and
further studies have indicated that the solutions' sets have
well-structured symmetry properties \cite{LU-Solutions}.

\section{Examples of the canonical form for three- and four-qubit state}

Here we give two simple examples of how we can get the canonical
forms of the arbitrary quantum state, and how we can verify whether
two quantum states in the canonical forms can be related by LU
symmetry $S$. As the entanglement classification of the mixed states
can be reduced to specific pure states case, here we only give
examples of pure states.

We randomly generate a $2\times 2\times 2$ pure state $\Psi$ with
the matrix unfolding
\begin{eqnarray}
\Psi_{(1)} = \begin{pmatrix}
 0.0260603 & 1.05491 & -3.69051 & 0.437711 \\
 1.25266 & 1.07259 & 3.2378 & 1.5625
\end{pmatrix} \; .
\end{eqnarray}
From the algorithm of Eq.(\ref{unfolding-singular}), the singular
value matrix $\Sigma$ is
\begin{eqnarray}
\begin{pmatrix}
 \sigma^{(1)}_1 &   \sigma^{(2)}_1 &  \sigma^{(3)}_1 \\
 \sigma^{(1)}_2 & \sigma^{(2)}_2 & \sigma^{(3)}_2
\end{pmatrix} =
\begin{pmatrix}
 5.03906 &  5.31586 & 5.17055\\ 2.27534 & 1.5202 & 1.95825
\end{pmatrix} \; .
\end{eqnarray}
The core tensor then is (unfolding with the first index)
\begin{eqnarray}
\Omega_{(1)} = \begin{pmatrix}
-5.01792 &  0.2815 & -0.354882 & -0.0862168 \\
0.19519 &  1.72088 & -1.17941 & -0.886923
\end{pmatrix} \; .
\end{eqnarray}

We give another example of four qubits state with degenerate
singular values. Two $2\times 2\times 2 \times 2$ quantum states
\begin{eqnarray}
\Psi_{(1)} & = & \frac{1}{\sqrt{10}}
\begin{pmatrix}
 1 & 0 & 0 & 0 & 0 & 0 & 2 & 0 \\
 0 & 1 & 0 & 0 & 0 & 0 & 0 & 2
\end{pmatrix} \; ,
\nonumber \\
\Psi_{(1)}' & = & \frac{1}{\sqrt{10}} \begin{pmatrix}
 1 & 0 & 0 & 0 & 0 & 0 & 2 & 0 \\
 0 & 1 & 0 & 0 & 0 & 0 & 0 & -2
\end{pmatrix} \; ,
\end{eqnarray}
are already the core tensors. The singular value matrices for them
are the same
\begin{eqnarray}
\begin{pmatrix}
 \sigma^{(1)}_1 &   \sigma^{(2)}_1 &  \sigma^{(3)}_1 & \sigma^{(4)}_1
 \\ \\
 \sigma^{(1)}_2 & \sigma^{(2)}_2 & \sigma^{(3)}_2 & \sigma^{(4)}_2
\end{pmatrix} =
\begin{pmatrix}
\frac{1}{2} &  \frac{4}{5} & \frac{4}{5} & \frac{1}{2}\\
\\ \frac{1}{2} & \frac{1}{5} & \frac{1}{5} & \frac{1}{2}
\end{pmatrix} = \begin{pmatrix}
 \sigma'^{(1)}_1 &   \sigma'^{(2)}_1 &  \sigma'^{(3)}_1 &
 \sigma'^{(4)}_1
 \\ \\
 \sigma'^{(1)}_2 & \sigma'^{(2)}_2 & \sigma'^{(3)}_2 &
 \sigma'^{(4)}_2
\end{pmatrix}\; .
\end{eqnarray}
In the vector forms of the matrices unfolding of $\Psi_{(1)}$ and
$\Psi'_{(1)}$, the symmetry $S$ takes the following form
\begin{eqnarray}
S = \begin{pmatrix} e^{i\theta_{1}^{(2)}+i\theta_{1}^{(3)}}
U^{(4)}\otimes U^{(1)} & 0 & 0 & 0
\\ 0 & e^{i\theta_{1}^{(2)}+i\theta_{2}^{(3)}} U^{(4)}\otimes U^{(1)} & 0
& 0  \\ 0 & 0 & e^{i\theta_{2}^{(2)}+i\theta_{1}^{(3)}}
U^{(4)}\otimes U^{(1)} & 0 \\ 0 & 0 & 0 &
e^{i\theta_{2}^{(2)}+i\theta_{2}^{(3)}}U^{(4)}\otimes U^{(1)}
\end{pmatrix} \; . \nonumber
\end{eqnarray}
The core tensors are then divided into four segments correspondingly
\begin{eqnarray}
\vec{\omega}_1 = \frac{1}{\sqrt{10}}\{ 1 , 0 , 0 , 1\}^{\mathrm{T}}
&  & \vec{\omega}'_1
=  \frac{1}{\sqrt{10}} \{ 1 , 0 , 0 , 1\}^{\mathrm{T}}  \; , \label{violation-1} \\
\vec{\omega}_2 = \frac{1}{\sqrt{10}} \{ 0 , 0 , 0 , 0\}^{\mathrm{T}}
& & \vec{\omega}_2' =  \frac{1}{\sqrt{10}} \{ 0 , 0 , 0 , 0\}^{\mathrm{T}}  \; , \\
\vec{\omega}_3 = \frac{1}{\sqrt{10}} \{ 0 , 0 , 0 , 0\}^{\mathrm{T}}
&  &
\vec{\omega}_3' = \frac{1}{\sqrt{10}} \{ 0 , 0 , 0 , 0\}^{\mathrm{T}}  \; , \\
\vec{\omega}_4 = \frac{1}{\sqrt{10}} \{ 2 , 0 , 0 , 2\}^{\mathrm{T}}
&  & \vec{\omega}_4' = \frac{1}{\sqrt{10}} \{ 2 , 0 , 0
,-2\}^{\mathrm{T}} \; , \label{violation-2}
\end{eqnarray}
Take all the above equations into
Eq.(\ref{detailed-symmetry-equation}) we have the following
effective equations:
\begin{eqnarray}
e^{i\theta_{1}^{(2)}+i\theta_{1}^{(3)}}U^{(4)}\otimes U^{(1)}
\vec{\omega}_1 = \vec{\omega}'_1 \; , \label{violation-solu1} \\
e^{i\theta_{2}^{(2)}+i\theta_{2}^{(3)}}U^{(4)}\otimes U^{(1)}
\vec{\omega}_4 = \vec{\omega}'_4 \; . \label{violation-solu2}
\end{eqnarray}
The components of $\vec{\omega}_1$, $\vec{\omega}_1'$ and
$\vec{\omega}_4$, $\vec{\omega}_4'$ in
Eqs.(\ref{violation-1},\ref{violation-2}) bring a contradiction to
Eqs.(\ref{violation-solu1},\ref{violation-solu2}). Now it is clear
that there is no solution for $U^{(4)}$ and $U^{(1)}$, and thus the
four qubits states $\Psi_{(1)}$ and $\Psi_{(1)}'$ are LU
inequivalent.

\section{Conclusions}

In summary, by using the tensor decomposition method we have
generalized the entanglement classification under LU equivalence to
arbitrary dimensional multipartite mixed states. The classification
actually reduces to the construction of the canonical forms of the
corresponding $N+1$-partite pure states. With the analysis of the
local symmetry in the canonical form, the core tensor can be
decomposed into a series of subtensors which are transformed
independently under the local symmetry. Base on this recognition of
the entanglement structure, a practical scheme is also developed for
the verification of local unitary equivalence of two multipartite
entangled states. In the verifications procedure, only in the worst
case of complete degeneracy for all the partite that we need to
solve multivariate polynomial equations. The well developed methods
and algorithms on solving such polynomial equations not only provide
the formula for finding the solutions but also impose well-formed
structures among the solutions \cite{LU-Solutions} which would shed
new light on the complete understanding of multipartite
entanglement.

\vspace{0.7cm} {\bf Acknowledgments}

This work was supported in part by the National Natural Science
Foundation of China(NSFC), by the CAS Key Projects KJCX2-yw-N29 and
H92A0200S2.

\end{document}